\documentclass{article}
\usepackage{graphicx}
\usepackage{natbib}
\usepackage{listings}
\usepackage{amsmath, amssymb}
\usepackage[a4paper, total={6in, 8in}]{geometry}
\usepackage{lineno}
\usepackage[colorlinks=true, allcolors=blue2]{hyperref}
\usepackage[T1]{fontenc}
\usepackage[utf8]{inputenc}
\usepackage{ebgaramond}
\usepackage{xcolor}
\usepackage{sectsty}  
\definecolor{blue2}{HTML}{12698A}
\usepackage[colorlinks=true, allcolors=blue2]{hyperref}
\sectionfont{\color{blue2}}
\subsectionfont{\color{blue2}}
\usepackage[font=small]{caption}  
\title{\bf  \color{blue2}Approximate Bayesian Computation Made Easy: A Practical Guide to ABC-SMC for Dynamical Systems with \texttt{pymc}}
\author{Mario Castro\\
Instituto de Investigaci\'on Tecnol\'ogica (IIT)\\
Grupo Interdisciplinar de Sistemas Complejos (GISC)\\
Universidad Pontificia Comillas, Madrid, Spain}
\date{\today}

\begin{document}
\maketitle

\begin{minipage}{0.6\textwidth}
\begin{abstract}
    Mechanistic models are essential tools across ecology, epidemiology, and the life sciences, but parameter inference remains challenging when likelihood functions are intractable. Approximate Bayesian Computation with Sequential Monte Carlo (ABC-SMC) offers a powerful likelihood-free alternative that requires only the ability to simulate data from mechanistic models. Despite its potential, many researchers remain hesitant to adopt these methods due to perceived complexity. This tutorial bridges that gap by providing a practical, example-driven introduction to ABC-SMC using Python. From predator-prey dynamics to hierarchical epidemic models, we illustrate by example how to implement, diagnose, and interpret ABC-SMC analyses. Each example builds intuition about when and why ABC-SMC works, how partial observability affects parameter identifiability, and how hierarchical structures naturally emerge in Bayesian frameworks. All code leverages PyMC's modern probabilistic programming interface, ensuring reproducibility and easy adaptation to new problems. The code its fully available for download at \href{https://github.com/mariocastro73/ABCSMC_pymc_by_example}{mariocastro73/ABCSMC\_pymc\_by\_example}
    \end{abstract}
\end{minipage}
\begin{minipage}{0.4\textwidth}
    \includegraphics[width=1\textwidth]{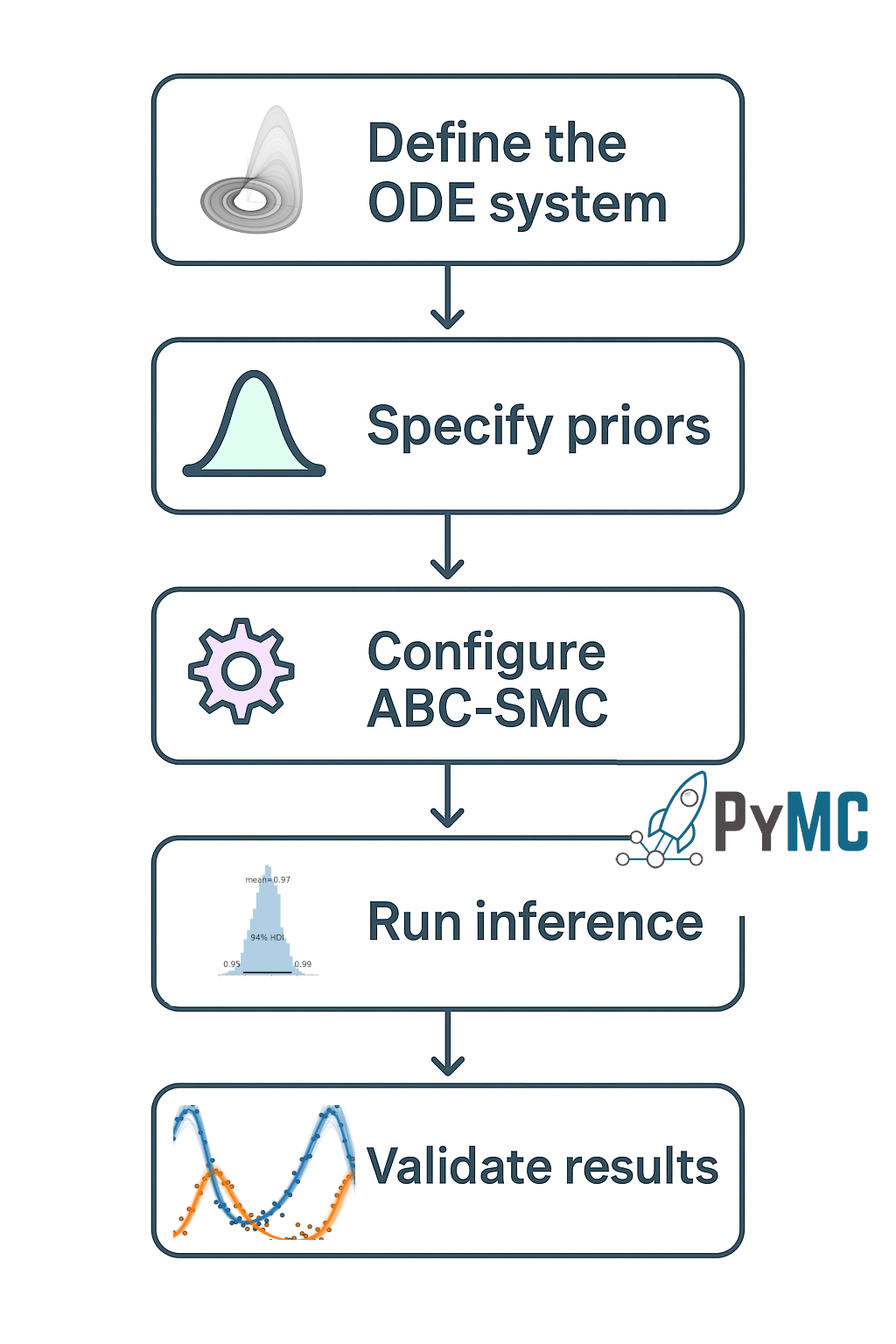}
\end{minipage}

\section{Introduction}
\label{sec:introduction}

Mechanistic models expressed through differential equations or stochastic processes are fundamental to understanding complex biological systems. These models formalize hypotheses, incorporate domain knowledge, and generate testable predictions about ecological and epidemiological dynamics. However, their practical utility depends on our ability to calibrate them against empirical data while properly accounting for uncertainty.

Bayesian inference provides an ideal framework for this task, yielding posterior distributions that quantify parameter uncertainty and facilitate model comparison. Yet many mechanistic models resist standard Bayesian approaches because their likelihood functions are analytically intractable, computationally prohibitive, or simply unavailable---particularly when systems contain latent variables, exhibit complex nonlinear dynamics, or involve stochastic processes.

Approximate Bayesian Computation (ABC) addresses precisely this challenge. By replacing likelihood evaluation with simulation and comparison, ABC makes Bayesian inference accessible for complex models where traditional methods fail \citep{Beaumont2010,Csillery2010,Sunnaker2013,toni2009approximate}. While early ABC implementations used simple rejection sampling, these approaches proved inefficient for realistic problems. The introduction of Sequential Monte Carlo (SMC) techniques dramatically improved computational efficiency, leading to the powerful ABC-SMC framework now widely used across scientific disciplines \citep{DelMoral2012,Bernardo2011}.

This tutorial aims to demystify ABC-SMC for researchers who build mechanistic models but lack extensive statistical training. Rather than focusing on theoretical details, we present four concrete examples that illustrate core concepts through practical implementation. Using Python and the PyMC library, we demonstrate how to:
\begin{itemize}
    \item Set up ABC-SMC analyses for ordinary differential equation models
    \item Interpret posterior distributions and diagnose convergence
    \item Reconstruct latent variables from partially observed systems
    \item Implement hierarchical models that share information across populations
\end{itemize}
\noindent
Our examples progress from simple to complex, building intuition at each step while maintaining technical reproducibility. By the end, readers will understand not just how to run ABC-SMC, but when to use it and how to interpret its results in the context of their own research questions.

All the code is available at \href{https://github.com/mariocastro73/ABCSMC_pymc_by_example}{mariocastro73/ABCSMC\_pymc\_by\_example} (it includes the code generating the figures).

\section{Bayesian inference without likelihoods: the ABC framework}
\label{sec:abc_framework}

\subsection{Why go Bayesian?}
Bayesian inference transforms prior knowledge about model parameters into updated posterior beliefs using observed data. Bayes' theorem formalizes this update:
\[
p(\theta \mid y) \propto p(y \mid \theta)\, p(\theta),
\]
where $p(\theta)$ represents prior knowledge about parameters $\theta$, $p(y \mid \theta)$ is the likelihood of observing data $y$ given those parameters, and $p(\theta \mid y)$ is the resulting posterior distribution.

This framework offers three critical advantages for dynamical systems:
\begin{enumerate}
    \item \textbf{Uncertainty quantification}: Rather than producing single-point parameter estimates, Bayesian methods characterize full uncertainty distributions, revealing which aspects of a model are well-constrained by data.
    \item \textbf{Integration of prior knowledge}: Domain expertise can inform plausible parameter ranges before seeing the data.
    \item \textbf{Model comparison}: Bayesian methods naturally accommodate formal comparison of alternative mechanistic hypotheses through Bayes factors or information criteria \citep{BDA3,Jaynes2003}.
\end{enumerate}
\noindent
For complex biological systems, these benefits are essential. However, they remain inaccessible when likelihood functions cannot be evaluated---a common situation in ecological and epidemiological modeling.

\subsection{The likelihood bottleneck in dynamical models}
Many realistic biological models resist likelihood-based inference for practical reasons:

\begin{itemize}
    \item \textbf{Latent variables}: In predator-prey systems, we might observe prey abundances but not predator populations; in epidemiology, exposed individuals often go undetected.
    \item \textbf{Complex noise structures}: Measurement errors may be non-Gaussian, heteroscedastic, or correlated in ways that defy simple characterization.
    \item \textbf{Computational barriers}: Stochastic simulation models (e.g., agent-based epidemic models) can generate realistic dynamics but lack closed-form likelihood expressions.
    \item \textbf{High-dimensional outputs}: Time-series data or spatial patterns contain rich information that's difficult to summarize in a tractable likelihood function.
\end{itemize}
\noindent
Traditional solutions---such as particle MCMC, data augmentation, or synthetic likelihoods---often require detailed noise models or computational resources beyond what many researchers can access \citep{Diggle1984,Price2018}. This reality motivates our focus on likelihood-free approaches that work directly with model simulations \citep{Cranmer2020}.

\subsection{ABC: Bypassing the likelihood}
Approximate Bayesian Computation sidesteps the likelihood evaluation problem through an elegant substitution: instead of calculating how probable the data are under given parameters, ABC assesses whether simulations from those parameters \textit{resemble} the observed data.

The fundamental ABC algorithm works as follows:
\begin{enumerate}
    \item Sample parameter values $\theta$ from the prior distribution $p(\theta)$
    \item Simulate synthetic data $x$ using the mechanistic model with parameters $\theta$
    \item Accept $\theta$ if the distance between simulated and observed data is small: $\rho(x,y) \leq \varepsilon$
\end{enumerate}
\noindent
Here, $\rho$ measures discrepancy between datasets (often Euclidean distance), and $\varepsilon$ controls approximation accuracy. As $\varepsilon \rightarrow 0$, the accepted samples converge to the true posterior distribution \citep{Beaumont2010,Csillery2010,Sunnaker2013}.

This simulation-based approach has profound practical implications: researchers can perform Bayesian inference using exactly the same simulators they've already built and validated, without deriving complex likelihood expressions. For scientists who think in terms of mechanistic processes rather than probability distributions, ABC aligns naturally with domain expertise.

\subsection{Practical considerations in ABC}
Three design choices critically impact ABC performance:
\begin{itemize}
    \item \textbf{Summary statistics}: When comparing high-dimensional data (e.g., time series), reducing dimensionality through informative summaries improves efficiency. Ideally, these should be sufficient statistics that preserve all information about parameters, though this is rarely achievable in practice. When possible, we recommend using the full data rather than summaries to avoid information loss.
    \item \textbf{Distance metrics}: Euclidean distance works well for many applications, but normalized or specialized distances might be needed when variables have different scales or units.
    \item \textbf{Tolerance schedules}: The acceptance threshold $\varepsilon$ balances computational efficiency against approximation accuracy. Small $\varepsilon$ yields precise posteriors but extremely low acceptance rates; large $\varepsilon$ runs quickly but produces crude approximations.
\end{itemize}
\noindent
These challenges motivated the development of more sophisticated ABC variants, particularly ABC-SMC \citep{toni2009approximate}, which we introduce next.

\section{Sequential Monte Carlo ABC (ABC--SMC)}
\label{sec:abc_smc}

\subsection{From rejection to sequential refinement}
Simple rejection ABC quickly becomes impractical as model complexity increases. Consider a typical scenario: a dynamical model with 5 parameters where only 1 in 10,000 random draws produces simulations within an acceptable distance of observed data. To obtain 500 posterior samples would require 5 million simulations---often computationally prohibitive.

ABC-SMC overcomes this limitation through sequential refinement. Instead of sampling directly from the prior, it constructs a series of intermediate distributions that gradually evolve toward the target posterior. Starting with a generous tolerance $\varepsilon_1$ that accepts many parameter values, the algorithm iteratively:
\begin{enumerate}
    \item Resamples promising parameter values from the current population
    \item Perturbs these values to explore nearby regions of parameter space
    \item Tightens the acceptance threshold to $\varepsilon_{t+1} < \varepsilon_t$
\end{enumerate}
\noindent
This staged approach concentrates computational effort where it matters most---in regions of parameter space consistent with the data---yielding orders-of-magnitude efficiency improvements over rejection sampling \citep{Toni2009,DelMoral2012}.

\subsection{The ABC-SMC algorithm}
Formally, ABC-SMC maintains a population of $N$ weighted particles $\{(\theta_t^{(i)}, w_t^{(i)})\}_{i=1}^N$ at each iteration $t$. The algorithm proceeds as follows:

\begin{enumerate}
    \item \textbf{Initialization}: Generate the first population using rejection ABC with large tolerance $\varepsilon_1$; assign equal weights $w_1^{(i)} = 1/N$
    \item \textbf{Iteration} (for $t = 2,\ldots,T$):
    \begin{enumerate}
        \item Decrease tolerance to $\varepsilon_t < \varepsilon_{t-1}$
        \item Resample $N$ particles from the previous population with probabilities proportional to their weights
        \item Perturb each resampled particle $\theta'$ using kernel $K_t(\theta \mid \theta')$
        \item Accept $\theta$ if $\rho(x,y) \leq \varepsilon_t$ where $x \sim p(x \mid \theta)$
        \item Assign weights $w_t^{(i)} \propto p(\theta)/\sum_j K_t(\theta \mid \theta_j')$
    \end{enumerate}
\end{enumerate}
\noindent
The perturbation kernel $K_t$ is typically a multivariate Gaussian with covariance tuned to the current particle population, ensuring efficient exploration while maintaining proximity to high-probability regions.

This adaptive structure provides crucial practical advantages:
\begin{itemize}
    \item Tolerances $\varepsilon_t$ can be selected automatically (e.g., as quantiles of previous distances)
    \item Resampling mitigates weight degeneracy while maintaining diversity
    \item Diagnostics like effective sample size (ESS) provide transparent convergence metrics
\end{itemize}

\subsection{Diagnostics for reliable inference}
Unlike MCMC methods, ABC-SMC doesn't produce dependent samples from a Markov chain. Instead, it generates weighted samples from a sequence of intermediate distributions. This difference necessitates specific diagnostic approaches:

\begin{itemize}
    \item \textbf{Tolerance trajectory}: A well-behaved run shows steady, monotonic decrease in tolerance values across iterations. Plateaus may indicate convergence to a stable posterior approximation.
    
    \item \textbf{Effective sample size (ESS)}: The ESS measures how many independent samples the weighted particle set represents. ESS values above 10\% of the total particle count typically indicate adequate population diversity.
    
    \item \textbf{Distance distributions}: As iterations progress, the distribution of distances $\rho(x,y)$ should shift toward zero while maintaining reasonable spread, indicating improved approximation without overfitting.
    
    \item \textbf{Posterior predictive checks}: Ultimately, the best diagnostic is whether simulations from posterior samples reproduce key features of the observed data.
\end{itemize}
\noindent
These diagnostics complement traditional Bayesian tools like trace plots and autocorrelation functions, which remain useful for assessing stability within each SMC iteration.

\subsection{Software landscape and implementation choices}
Several Python libraries implement ABC-SMC, each with distinct strengths:
\begin{itemize}
    \item \texttt{ABC-SysBio} pioneered likelihood-free inference for systems biology but is no longer actively maintained \citep{Liepe2010}
    \item \texttt{pyABC} offers specialized ABC-SMC functionality with excellent scalability for large problems \citep{Klinger2018}
    \item \texttt{ELFI} extends beyond ABC to include synthetic likelihood and regression adjustment methods \citep{Lintusaari2018}
    \item Modern libraries like \texttt{sbi} leverage neural networks for high-dimensional inference problems \citep{Tejero2020}
\end{itemize}
\noindent
For this tutorial, we use \texttt{PyMC} because it uniquely bridges likelihood-free and likelihood-based approaches within a single framework \citep{Salvatier2016,PyMC2023}. PyMC's \texttt{pm.Simulator} interface wraps any generative model for ABC while maintaining compatibility with Hamiltonian Monte Carlo, variational inference, and other Bayesian methods~\cite{BDA3}. This unified approach lets researchers:
\begin{itemize}
    \item Start with intuitive ABC methods and gradually incorporate likelihood-based techniques
    \item Leverage consistent syntax for model specification, diagnostics, and visualization
    \item Benefit from automatic integration with ArviZ for posterior analysis \citep{Kumar2019}
    \item Scale efficiently using modern computational backends (JAX, Aesara)
\end{itemize}
\noindent
By building all examples around PyMC, we provide a coherent pathway from simulation-based inference to more advanced Bayesian techniques without requiring readers to learn multiple specialized frameworks.

\section{Case studies}
\label{sec:case_studies}

We now demonstrate ABC-SMC through four progressively complex examples. Each case study introduces new concepts while reinforcing core principles, creating an accessible learning path for practitioners. All code examples use PyMC's implementation of ABC-SMC and follow identical workflow patterns: model specification, prior definition, inference execution, and posterior validation.

\subsection{Lotka--Volterra with two observed species}
\label{sec:lv_two_species}

The Lotka--Volterra predator-prey model provides an ideal starting point. Its dynamics follow:
\begin{equation}
  \frac{\mathrm{d}X_1}{\mathrm{d}t} = a X_1 - b X_1 X_2, \quad
  \frac{\mathrm{d}X_2}{\mathrm{d}t} = -c X_2 + d b X_1 X_2,
\end{equation}
\noindent
where $X_1$ represents prey, $X_2$ predators, $a$ is prey growth rate, $b$ is predation rate, $c$ is predator mortality, and $d$ is conversion efficiency. We assume $c$ and $d$ are known, focusing inference on $a$ and $b$.

In this first example, we observe both species across time, creating an identifiable system where parameters should be precisely constrained. The implementation follows these steps:

\begin{enumerate}
  \item Define the ODE system and wrap it as a simulator function
  \item Specify weakly informative prior distributions for unknown parameters
  \item Configure ABC-SMC with an initial tolerance and number of particles
  \item Run inference and analyze the resulting posterior distribution
  \item Validate results through posterior predictive checks
\end{enumerate}

\noindent
The simulator function (Figure~\ref{lst:lv-ode}) numerically solves the ODE system and returns flattened trajectories suitable for distance calculation in PyMC.

\begin{figure}[!htp]
\begin{lstlisting}
c = 1.5
d = 0.75
# Observed  time points and populations
data = np.loadtxt("observed_data.csv", delimiter=",", skiprows=1)
t = data[:, 0]
X0 = data[1, 1:3]                 # initial prey and predator
observed_matrix = data[:, 1:3]
observed = observed_matrix.reshape(-1)
def dX_dt(X, t, a, b, c, d):
    return np.array([a * X[0] - b * X[0] * X[1],
                     -c * X[1] + d * b * X[0] * X[1]])
def competition_model(rng, a, b, size=None):
    a_scalar = float(a)
    b_scalar = float(b)
    result = odeint(dX_dt, y0=X0, t=t, rtol=0.01,
                    args=(a_scalar, b_scalar, c, d))
    return result.reshape(-1)
\end{lstlisting}
 \caption{Mechanistic Lotka--Volterra model and simulator.  
  The function \texttt{competition\_model} maps parameters $(a,b)$ to a simulated prey--predator
  trajectory evaluated at the observation times, flattened into a one-dimensional array suitable
  for use in \texttt{pm.Simulator}.}
  \label{lst:lv-ode}
  \end{figure}

\noindent
Within the PyMC model context (Figure~\ref{lst:lv-pymc}), we place weakly informative half-normal priors on $a$ and $b$, then wrap the simulator using \texttt{pm.Simulator}. This interface handles the distance calculations and adaptive tolerance scheduling automatically.

\begin{figure}[!htp]
\begin{lstlisting}
with pm.Model() as model_lv:
    # Priors on growth and predation rates
    a = pm.HalfNormal("a", sigma=1.0)
    b = pm.HalfNormal("b", sigma=1.0)
    # Likelihood-free observation model (ABC)
    sim = pm.Simulator(
        "sim",
        competition_model,
        params=(a, b),
        epsilon=10.0,          # initial tolerance
        observed=observed,
    )
    # ABC-SMC inference
    samples = pm.sample_smc(draws=500, chains=4)
    posterior = samples.posterior.stack(samples=("draw", "chain"))
  
\end{lstlisting}
\caption{\label{lst:lv-pymc}PyMC model for ABC--SMC inference in the Lotka--Volterra system.  
  The \texttt{pm.Simulator} distribution wraps the ODE-based simulator and uses an initial tolerance
  \texttt{epsilon=10.0}.  The call to \texttt{pm.sample\_smc} runs the ABC--SMC algorithm and returns
  an \texttt{InferenceData} object containing the approximate posterior samples.}
\end{figure}

\begin{figure}[!htp]
  \centering
  \begin{tabular}{ll}
       (A)& (B)  \\
  \includegraphics[width=0.6\textwidth]{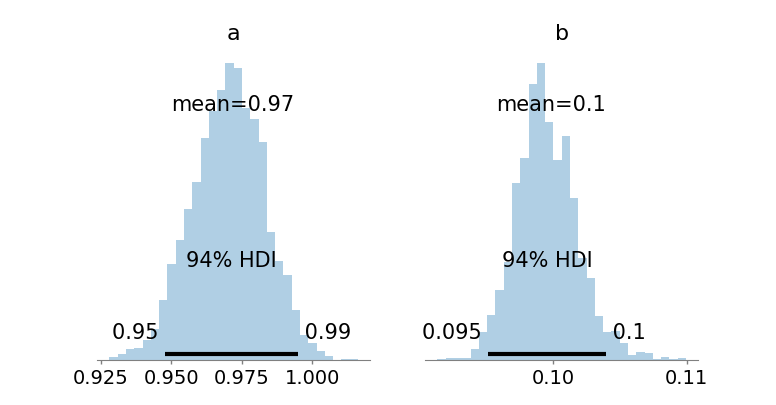}&
  \includegraphics[width=0.4\textwidth]{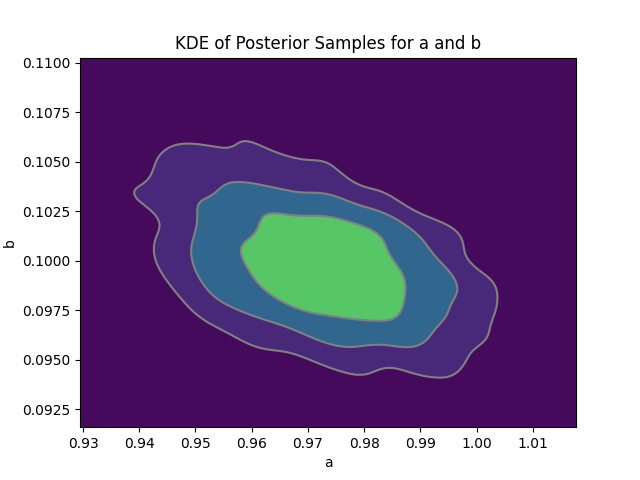}
  \end{tabular}
  \caption{(Left) Marginal posterior distributions of the prey growth rate $a$ and predation rate $b$
  obtained by ABC--SMC.  
  The histograms and shaded regions show the approximate posterior densities together with
  high-density intervals.  
  Both parameters are well-identified, with posterior mass concentrated around the true generating
  values used in the synthetic data. \textbf{(Right)} Joint posterior distribution of $(a,b)$, visualized via kernel density estimation.
  The nested contours represent regions of increasing posterior probability density.
  The elongated shape of the contours reveals a correlation between $a$ and $b$, indicating that
  certain combinations of growth and predation rates are jointly plausible given the data.}
  \label{fig:lv-posteriors}
\end{figure}

\noindent
The resulting posterior distributions (Figure~\ref{fig:lv-posteriors}) reveal precisely identified parameters with unimodal distributions centered near their true values. The joint posterior shows a negative correlation between $a$ and $b$, reflecting how increased prey growth can be compensated by higher predation rates while maintaining similar dynamics.

\begin{figure}[!htp]
  \centering
  \includegraphics[width=\textwidth]{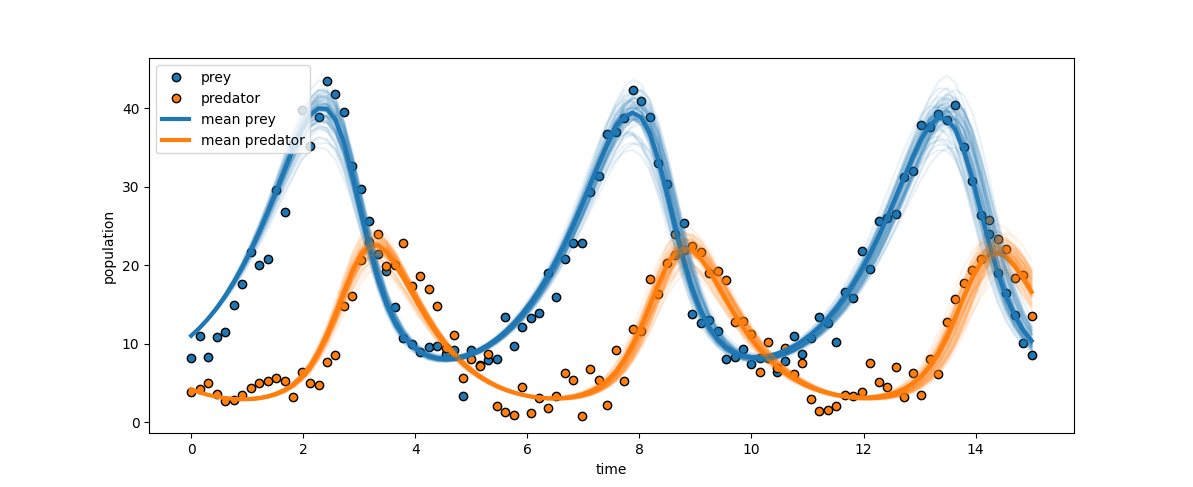}
  \caption{Posterior predictive trajectories for the Lotka--Volterra system with both species observed.
  The points denote the observed prey (blue) and predator (orange) time series.  
  The thick lines show the trajectories at the posterior mean parameters, while the light curves
  represent simulations from randomly selected posterior samples.  
  The close alignment between the observed data and the posterior predictive ensemble indicates that
  the inferred parameters provide a good description of the predator--prey dynamics.}
  \label{fig:lv-posterior-predictive}
\end{figure}

Posterior predictive checks (Figure~\ref{fig:lv-posterior-predictive}) confirm the model's ability to reproduce observed oscillations. We compute trajectories at the posterior mean parameters and overlay simulations from randomly selected posterior samples, creating an uncertainty envelope that tightly encompasses the observed data. This visualization demonstrates how Bayesian inference quantifies not just parameter uncertainty, but its propagation to model predictions.

\subsection{Lotka--Volterra with one unobserved species}
\label{sec:lv_one_unobserved}

Real-world ecological monitoring rarely captures all interacting species with equal fidelity. In our second case study, we consider a more realistic scenario where only prey populations are observed, while predator dynamics remain latent.

The computational implementation requires only one modification: our simulator now returns only the prey trajectory for comparison with observations (Figure~\ref{lst:lv-competition-unobserved}). Crucially, the full predator-prey dynamics still constrain inference---parameter combinations that produce plausible prey cycles but biologically unreasonable predator trajectories will be rejected during ABC-SMC sampling.
\begin{figure}[!htp]
\begin{lstlisting}
# Only the first species (prey) is observed
observed_vector = data[:, 1]
observed = observed_vector
def competition_model(rng, a, b, size=None):
    a_scalar = a.item() if hasattr(a, "item") else float(a)
    b_scalar = b.item() if hasattr(b, "item") else float(b)
    result = odeint(dX_dt, y0=X0, t=t, rtol=0.01,
                    args=(a_scalar, b_scalar, c, d))
    # Return only the prey trajectory; the predator is latent
    return result[:, 0]
  \end{lstlisting}
  \caption{Simulator for the partially observed Lotka--Volterra model.  
  In contrast to the fully observed case (where the flattened prey--predator matrix was
  returned), the distance in ABC is now computed only on the prey trajectory, while the
  predator dynamics remain latent but constrained through the ODE system.}
  \label{lst:lv-competition-unobserved}
\end{figure}
\noindent
This partial observability fundamentally changes the inference problem. While the prey time series still contain information about predator dynamics through their oscillatory structure, many more parameter combinations can produce similar prey trajectories with different underlying predator dynamics.

\begin{figure}[!htp]
  \centering
  \includegraphics[width=\textwidth]{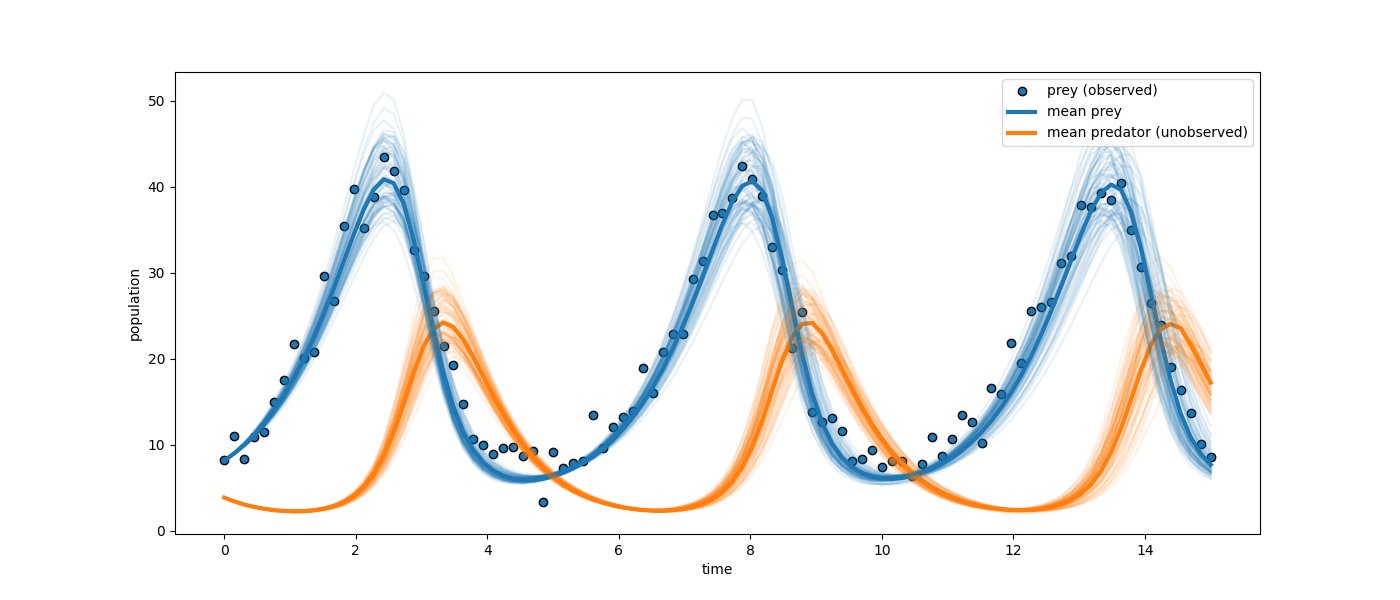}
  \caption{Posterior predictive trajectories for the Lotka--Volterra system when only the prey
  population is observed.  Blue points show the observed prey time series, and the thick blue line
  is the prey trajectory at the posterior mean parameters.  The orange curves represent the
  corresponding posterior predictive predator trajectories, which are entirely latent in the data.
  The ensemble of semi-transparent curves visualizes uncertainty: the prey dynamics are tightly
  constrained by the observations, whereas the predator trajectories exhibit larger variability,
  yet remain dynamically consistent with the observed prey cycles.}
  \label{fig:lv-unobserved}
\end{figure}

The posterior predictive results (Figure~\ref{fig:lv-unobserved}) illustrate this uncertainty clearly: prey trajectories remain tightly constrained around observations, but predator trajectories display substantially wider uncertainty envelopes. Despite this increased uncertainty, the model successfully reconstructs the qualitative features of predator dynamics---including oscillation phase and approximate amplitude---demonstrating how mechanistic constraints can compensate for incomplete observations.

\subsection{SEIR model with infected data only}
\label{sec:seir_singlepop}

Epidemiological models present another domain where ABC-SMC excels. Compartmental models like SEIR (Susceptible-Exposed-Infected-Recovered) describe disease spread through populations with partially observed states. Our third example considers a standard SEIR model with dynamics:

\begin{align}
  \frac{\mathrm{d}S}{\mathrm{d}t} &= -\beta \frac{S I}{N}, &
  \frac{\mathrm{d}E}{\mathrm{d}t} &= \beta \frac{S I}{N} - \sigma E, \\
  \frac{\mathrm{d}I}{\mathrm{d}t} &= \sigma E - \gamma I, &
  \frac{\mathrm{d}R}{\mathrm{d}t} &= \gamma I,
\end{align}

\noindent
where $\beta$ is the transmission rate, $\sigma$ is the incubation rate (progression from exposed to infected), and $\gamma$ is the recovery rate. We observe only $S$, $I$, and $R$ compartments, while the exposed compartment $E$ remains latent (Figure~\ref{lst:seir-simulator}).

\begin{figure}[!htp]
\begin{lstlisting}
# Observed  columns S, I, R (no E)
data = np.loadtxt("epidemic_data.csv", delimiter=",", skiprows=1)
t = data[:, 0]
X0 = [data[0, 1], 0, data[0, 2], data[0, 3]]
observed_matrix = data[:, 1:4]      # S, I, R
observed = observed_matrix.reshape(-1)
def dX_dt(X, t, beta, sigma, gamma):
    S, E, I, R = X
    N = S + E + I + R
    dS = -beta * S * I / N
    dE = beta * S * I / N - sigma * E
    dI = sigma * E - gamma * I
    dR = gamma * I
    return np.array([dS, dE, dI, dR])
    
def seir_model(rng, beta, sigma, gamma, size=None):
    beta_scalar = beta.item() if hasattr(beta, "item") else float(beta)
    sigma_scalar = sigma.item() if hasattr(sigma, "item") else float(sigma)
    gamma_scalar = gamma.item() if hasattr(gamma, "item") else float(gamma)
    result = odeint(dX_dt, y0=X0, t=t, rtol=0.01,
                    args=(beta_scalar, sigma_scalar, gamma_scalar))
    # Return only S, I, R (columns 0, 2, 3)
    return result[:, [0, 2, 3]].reshape(-1)
 
\end{lstlisting}
 \caption{Mechanistic SEIR model and simulator.  
  The function \texttt{seir\_model} maps parameters $(\beta,\sigma,\gamma)$ to simulated
  trajectories for $S$, $I$, and $R$, which are flattened into a vector suitable for
  \texttt{pm.Simulator}.  The exposed compartment $E$ is latent and inferred indirectly
  through its impact on the infected and recovered dynamics.}
  \label{lst:seir-simulator}
\end{figure}

\noindent
Within the PyMC framework (Figure~\ref{lst:seir-pymc}), we place weakly informative priors on epidemiological parameters and perform ABC-SMC inference following the same pattern as previous examples.

\begin{figure}[!htp]
\begin{lstlisting}
with pm.Model() as model_seir:
    # Priors on transmission, incubation, and recovery rates
    beta  = pm.HalfNormal("beta",  2.0)
    sigma = pm.HalfNormal("sigma", 1.0)
    gamma = pm.HalfNormal("gamma", 0.5)
    # Likelihood-free observation model (ABC)
    sim = pm.Simulator(
        "sim",
        seir_model,
        params=(beta, sigma, gamma),
        epsilon=0.5,
        observed=observed,
    )
    # ABC-SMC inference
    samples = pm.sample_smc(draws=500, chains=4)
    posterior = samples.posterior.stack(samples=("draw", "chain"))
  \end{lstlisting}
  \caption{PyMC model for ABC--SMC inference in the SEIR system.  
  The \texttt{pm.Simulator} distribution wraps the SEIR simulator, and the call to
  \texttt{pm.sample\_smc} produces posterior samples for $(\beta,\sigma,\gamma)$ and
  corresponding latent trajectories.}
  \label{lst:seir-pymc}
\end{figure}
\begin{figure}[!htp]
  \centering
  \includegraphics[width=0.8\textwidth]{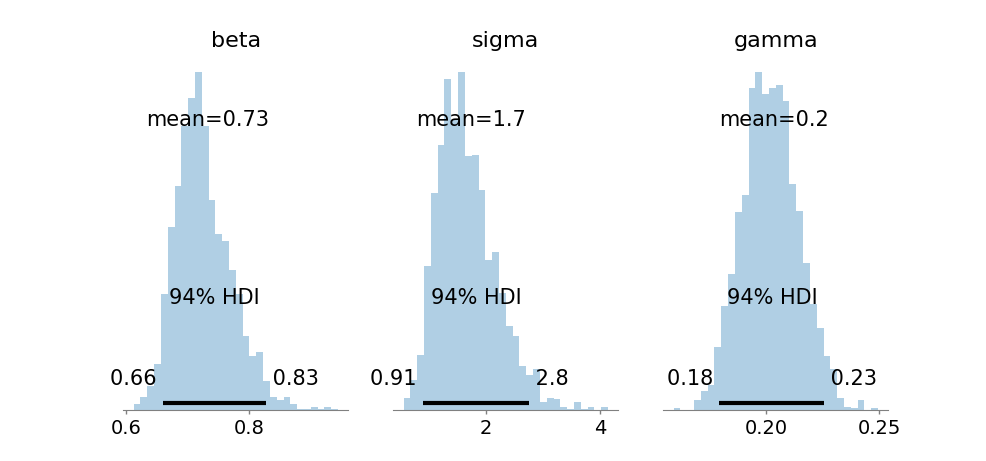}
  \caption{Marginal posterior distributions of the SEIR parameters $\beta$, $\sigma$, and
  $\gamma$ obtained by ABC--SMC.  The histograms and high-density intervals (HDIs)
  quantify the uncertainty in transmission, incubation, and recovery rates given the
  observed $(S,I,R)$ time series.}
  \label{fig:seir-posteriors}
\end{figure}
\begin{figure}[!htp]
  \centering
  \includegraphics[width=0.7\textwidth]{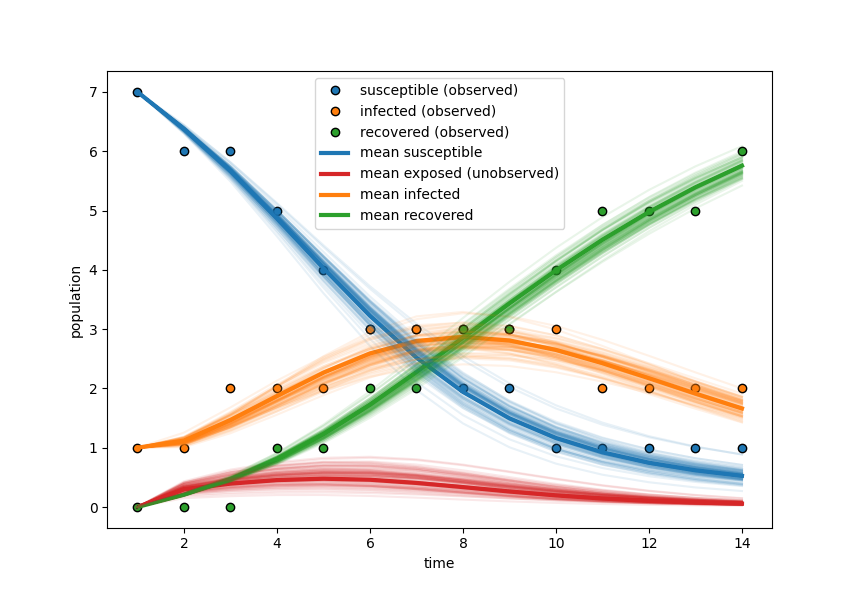}
  \caption{Posterior predictive trajectories for the SEIR model.  
  Points denote observed susceptible (blue), infected (orange), and recovered (green)
  counts.  Thick lines show the mean trajectories for $S$, $E$ (latent), $I$, and $R$
  under the posterior, and the light curves correspond to simulations from randomly
  selected posterior samples.  The exposed compartment $E$ is never directly observed,
  yet its dynamics are reconstructed from the information carried by the observed
  compartments.}
  \label{fig:seir-posterior-predictive}
\end{figure}

\noindent
The posterior distributions (Figure~\ref{fig:seir-posteriors}) show that all three epidemiological parameters are reasonably well identified despite the unobserved exposed compartment. The posterior predictive trajectories (Figure~\ref{fig:seir-posterior-predictive}) demonstrate how the model reconstructs latent dynamics: while the exposed compartment shows greater uncertainty than observed compartments, its characteristic peak timing and magnitude remain well-constrained by the interlocking dynamics of the full system.

\subsection{Hierarchical SEIR with two groups}
\label{sec:seir_hierarchical}

Our final example demonstrates how ABC-SMC naturally extends to hierarchical models that share information across related populations. This structure is essential when analyzing heterogeneous systems like multi-city epidemics or age-structured disease spread.

Consider two subpopulations experiencing the same epidemic process but with potentially different transmission dynamics. In a hierarchical model:
\begin{itemize}
    \item Each group has its own parameters $(\beta_g, \sigma_g, \gamma_g)$
    \item These group-level parameters are drawn from shared population distributions with hyperparameters $(\mu_\beta, \sigma_\beta, \mu_\sigma, \sigma_\sigma, \mu_\gamma, \sigma_\gamma)$
    \item The data from all groups jointly inform both group-specific and population-level parameters
\end{itemize}
\noindent
This structure enables "borrowing strength" across groups: populations with sparse data have their estimates stabilized by information from better-observed groups, while still allowing for meaningful heterogeneity.

\begin{figure}[!htp]
  \centering
  \includegraphics[width=\textwidth]{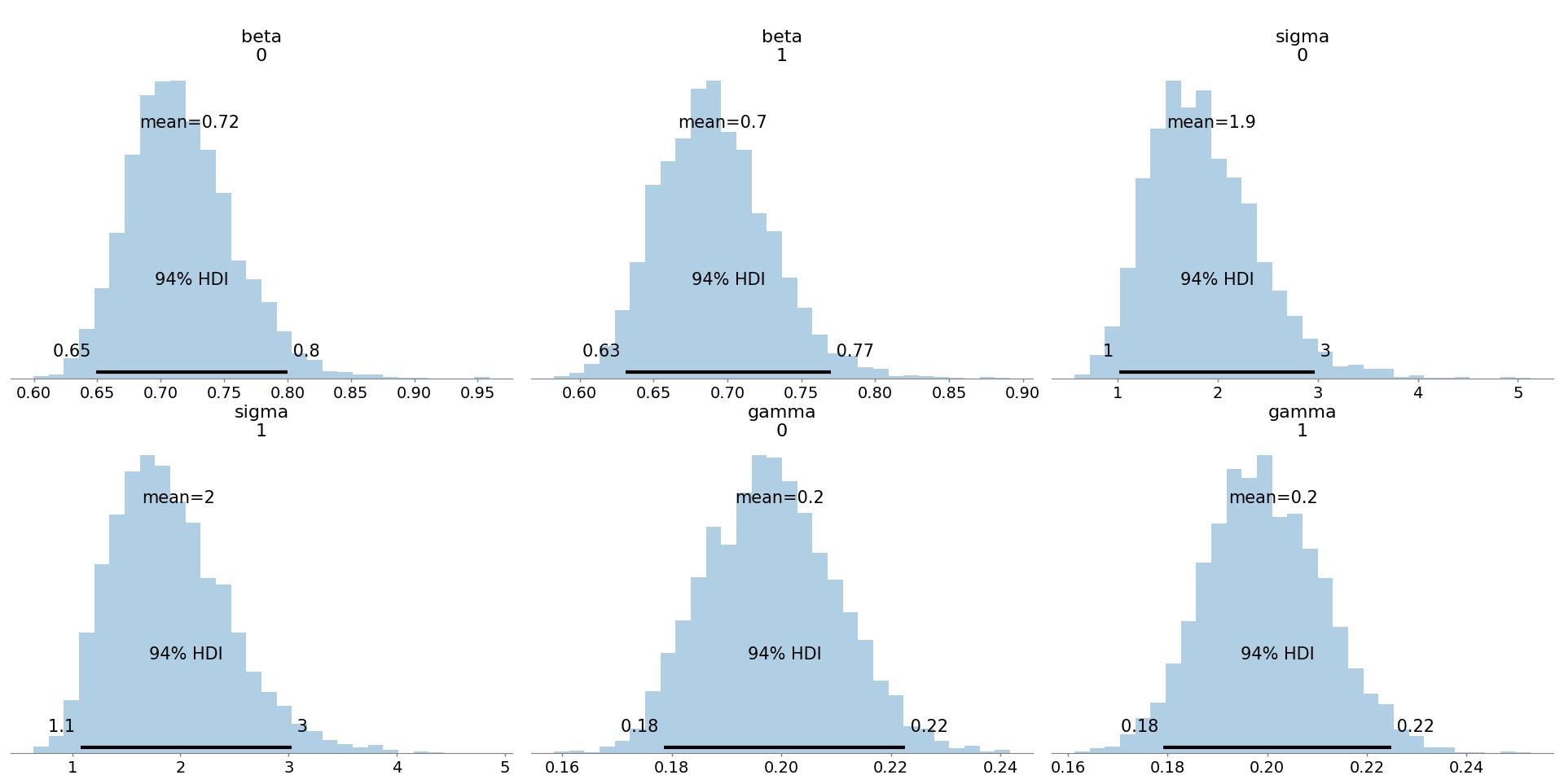}
  \caption{Posterior marginal distributions of group-specific parameters in the hierarchical SEIR
  model.  
  Each row corresponds to a parameter: transmission rate $\beta$, incubation rate $\sigma$, and
  recovery rate $\gamma$.  
  Columns correspond to the two groups.  
  The broader distributions compared to the single-population model reflect both cross-group
  heterogeneity and partial pooling through the hierarchical structure.}
  \label{fig:seir-hierarchical-marginals}
\end{figure}

\begin{figure}[!htp]
  \centering
  \includegraphics[width=\textwidth]{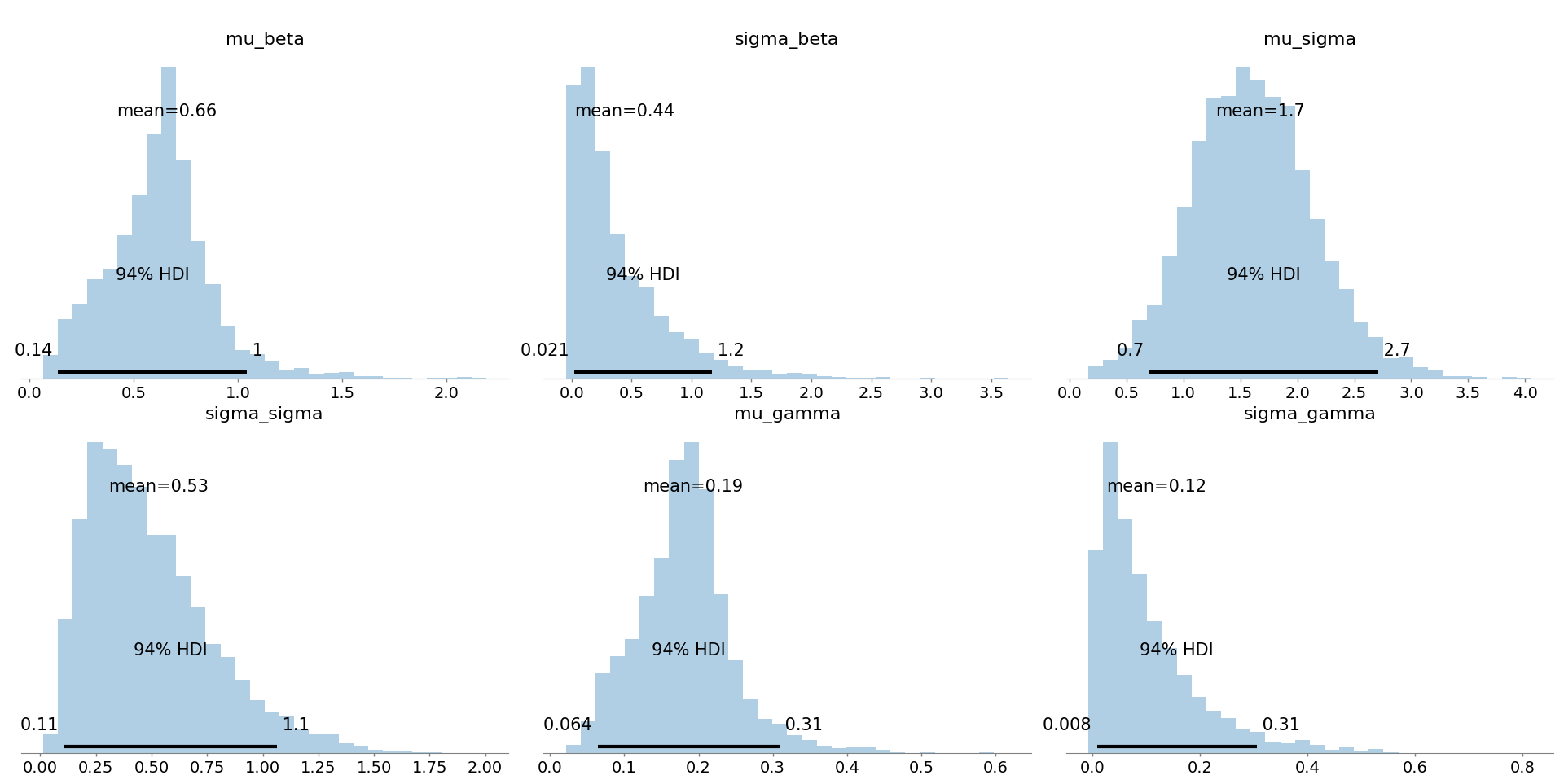}
  \caption{Posterior distributions of the hyperparameters in the hierarchical SEIR model.
  The means $(\mu_\beta, \mu_\sigma, \mu_\gamma)$ describe the population-level average
  transmission, incubation, and recovery rates, while the scales
  $(\sigma_\beta, \sigma_\sigma, \sigma_\gamma)$ quantify variability between groups.
  Strong pooling corresponds to small scales; weak pooling corresponds to large scales.
  This structure clarifies how individual group parameters relate to population-level
  epidemiological dynamics.}
  \label{fig:seir-hyperparameters}
\end{figure}

Implementing this model requires extending our simulator to handle multiple populations simultaneously and specifying hierarchical prior structures. The resulting posterior (Figure~\ref{fig:seir-hierarchical-marginals}) shows wider distributions for group-specific parameters compared to the single-population model, reflecting the additional uncertainty from partial pooling. The hyperparameter posteriors (Figure~\ref{fig:seir-hyperparameters}) reveal the population-level patterns: for example, recovery rates ($\gamma$) show less between-group variation than incubation rates ($\sigma$), suggesting more consistent biological processes across populations for recovery than for disease progression.

\noindent
Hierarchical models provide two critical advantages over analyzing groups separately:
\begin{enumerate}
    \item \textbf{Stability for data-limited groups}: When one population has sparse observations, its parameter estimates are stabilized by information from better-observed groups, reducing overfitting.
    \item \textbf{Quantification of heterogeneity}: Rather than assuming either identical parameters across groups or completely independent estimates, hierarchical models objectively assess how much parameters truly vary between populations.
\end{enumerate}
\noindent
These benefits make hierarchical ABC-SMC particularly valuable for real-world applications where data quality and quantity vary substantially across observational units.

\section{Practical guidance and best practices}
\label{sec:guidance}

Our examples illustrate both the capabilities and limitations of ABC-SMC. Based on this experience, we offer concrete guidance for practitioners:

\subsection{When to choose ABC-SMC}
ABC-SMC excels in scenarios where:
\begin{itemize}
    \item The generative model is well-understood but the likelihood is intractable
    \item Systems contain latent variables that cannot be directly observed
    \item Model outputs are high-dimensional (e.g., time series, spatial patterns)
    \item Prior knowledge exists in the form of parameter ranges rather than precise distributions
\end{itemize}
\noindent
Likelihood-based approaches remain preferable when the likelihood is tractable and differentiable, as they typically provide more precise inference with less computational effort.

\subsection{Implementation recommendations}
\begin{itemize}
    \item \textbf{Full data over summaries}: When computationally feasible, use complete time-series data rather than summary statistics to preserve information
    
    \item \textbf{Appropriate distance metrics}: For variables spanning multiple orders of magnitude, compute distances on log-transformed values; when shape matters more than absolute values, consider normalized distances
    
    \item \textbf{Informative priors}: Weakly informative priors dramatically improve efficiency by focusing computation on plausible regions; avoid overly broad priors that waste simulations on biologically impossible parameter combinations
    
    \item \textbf{Adaptive tolerances}: Let the algorithm determine tolerance schedules automatically based on distance quantiles rather than setting fixed values
    
    \item \textbf{Adequate particle populations}: Use sufficient particles to maintain population diversity (typically 500-2000); monitor effective sample size to detect degeneracy
\end{itemize}

\subsection{Diagnostics and validation}
Always perform these critical checks:
\begin{itemize}
    \item \textbf{Posterior predictive validation}: Simulate data from posterior parameter samples and verify they reproduce key features of observed data
    \item \textbf{Tolerance trajectory monitoring}: Ensure tolerances decrease smoothly without premature plateaus
    \item \textbf{Sensitivity analysis}: Test how results change with different prior specifications, distance metrics, and tolerance schedules
    \item \textbf{Model comparison}: When multiple mechanistic hypotheses exist, use ABC model selection approaches to objectively compare alternatives
\end{itemize}
\noindent
Most importantly, communicate uncertainty explicitly: present full posterior distributions rather than point estimates, and show posterior predictive envelopes rather than single best-fit trajectories.

\section{Discussion and conclusions}
\label{sec:conclusions}

This tutorial has demonstrated how ABC-SMC bridges the gap between mechanistic modeling and rigorous statistical inference for complex biological systems. By replacing intractable likelihood calculations with simulation and comparison, ABC-SMC makes Bayesian inference accessible for models that were previously resistant to formal statistical analysis.

Our four case studies illustrate key principles that apply across domains:
\begin{itemize}
    \item \textbf{Partial observability creates uncertainty}: Unobserved variables (predators, exposed individuals) increase parameter uncertainty but can often be reconstructed through dynamical constraints
    
    \item \textbf{Structure informs identifiability}: Even with limited data, mechanistic model structure constrains plausible parameter combinations and latent trajectories \citep{castro2020testing,villaverde2016structural}
    
    \item \textbf{Hierarchical models share information}: Population-level patterns stabilize estimates for data-limited groups while quantifying meaningful heterogeneity
    
    \item \textbf{Uncertainty propagation matters}: Bayesian methods naturally propagate uncertainty from parameters to predictions, providing honest assessments of model reliability
\end{itemize}
\noindent
The PyMC implementation we've demonstrated offers particular advantages for scientists new to Bayesian methods. Its unified interface handles both likelihood-based and likelihood-free inference, allowing researchers to start with intuitive ABC methods and gradually incorporate more advanced techniques as needed.

All code and data for this tutorial are available in the GitHub repository  \href{https://github.com/mariocastro73/ABCSMC_pymc_by_example}{mariocastro73/ABCSMC\_pymc\_by\_example}, which includes complete, executable notebooks for each case study. We encourage readers to experiment with these examples---modifying priors, adjusting tolerances, or introducing model variants---to build intuition for their own research applications.

As computational resources continue to expand and algorithms improve, simulation-based Bayesian inference will become increasingly accessible for complex biological models. By mastering these techniques now, researchers can extract maximum insight from their mechanistic models while maintaining statistical rigor---ultimately leading to more reliable scientific conclusions and better-informed decisions in ecology, epidemiology, and beyond.

\section*{Acknowledgments}
The author thanks the PyMC development team for creating and maintaining an exceptional open-source library, and acknowledges helpful feedback from colleagues at the Institute for Technological Research. This work was supported by the Spanish Ministry of Science and the Madrid Regional Government.

\bibliographystyle{plainnat}
\bibliography{refs}

\end{document}